\documentclass[prd,aps,tightenlines,preprint,showpacs,showkeys]{revtex4}

\usepackage{graphicx}
\usepackage{bm}

\newcommand{\OO}[1]{{\mathcal O}(c^{-#1})}
\newcommand{\ve}[1]{\bm{#1}}
\newcommand{\muas}{\hbox{$\mu$as}}
\newcommand{\arcsec}{\hbox{$^{\prime\prime}$}}

\newcommand{\G}{{\mathcal G}}
\newcommand{\W}{{\mathcal W}}
\newcommand{\T}{{\mathcal T}}
\newcommand{\X}{{\mathcal X}}
\newcommand{\A}{{\mathcal A}}
\newcommand{\B}{{\mathcal B}}
\newcommand{\C}{{\mathcal C}}
\newcommand{\D}{{\mathcal D}}
\newcommand{\R}{{\mathcal R}}
\renewcommand{\P}{{\mathcal P}}
\newcommand{\Q}{{\mathcal Q}}
\renewcommand{\P}{{\mathcal P}}
\renewcommand{\L}{{\mathcal L}}
\renewcommand{\S}{{\mathcal S}}
\renewcommand{\o}{{\rm o}}

\begin{document}

\title{Physically adequate proper reference system of a test observer
and relativistic description of the GAIA attitude}

\author{Sergei A. \surname{Klioner}}

\affiliation{Lohrmann Observatory, Dresden Technical University,
Mommsenstr. 13, 01062 Dresden, Germany}


\date{\today}

\begin{abstract}
A relativistic definition of the physically adequate proper reference
system of a test observer is suggested within the framework of the PPN
formalism. According to the nomenclature accepted within the GAIA
project this reference system is called Center-of-Mass Reference System
(CoMRS). The interrelation between the suggested definition of
the CoMRS and the Resolutions
2000 on relativity of the International Astronomical Union (IAU)
are elucidated. The tetrad representation of the
CoMRS at its origin is also explicated. It is demonstrated how to use
that tetrad representation to calculate the relation between the
observed direction of a light ray and the corresponding coordinate
direction in the Barycentric Celestial Reference System of the IAU. It
is argued that the kinematically non-rotating CoMRS is the natural
choice of the reference system where the attitude of the observer (e.g.
of the GAIA satellite) should be modeled. The relativistic equations of
rotational motion of a satellite relative to its CoMRS are briefly
discussed. A simple algorithm for the attitude description of the
satellite is proposed.
\end{abstract}

\keywords{general relativity, reference systems, astrometry, GAIA}
\pacs{04.25.Nx,04.80.Cc,95.10.Ce}

\maketitle

\section{Introduction}
\label{Section-Introduction}

Future space astrometry projects like GAIA
\cite{GAIA:2000,Perryman:et:al:2001,Bienayme:Turon:2002} and SIM \cite
{SIM:1998} are expected to attain an accuracy of 1 microarcsecond
(\muas) for positions of remote celestial sources. This high accuracy
requires general relativity to be used for data modeling. A
relativistic model of positional observation with microarcsecond
accuracy involves many subtle details. Recently a number of such models
have been suggested (see \cite{Klioner:Kopeikin:1992,Klioner:2003,
Bini:deFelice:2003,Bini:Crosta:deFelice:2003} and references therein).
The purpose of this paper is, first of all, to provide a relativistic
definition of the physically adequate local (proper) reference system
of a test observer (e.g., a satellite). According to the nomenclature
adopted within the GAIA project \cite{Bastian:2003} this reference
system is called Center-of-Mass Reference System (CoMRS) below. As a
co-product, the calculation of the observed light direction as adopted
in \cite{Klioner:2003} is explained in detail and explicitly justified.
The CoMRS is intended to be used to describe physical phenomena located
within the immediate vicinity of the observer (e.g., the rotational
motion of the satellite, the process of observation, etc.). This
reference system can be used to define the attitude parameters of the
satellite. In order to define the CoMRS we make use of the parametrized
post-Newtonian (PPN) version of the relativistic framework adopted
recently by the International Astronomical Union (IAU) for the use for
high-precision astrometry, celestial mechanics, geodesy and metrology
\cite{Soffel:et:al:2003}. The IAU Resolution B1.3 (2000) adopted at the
\expandafter\uppercase\expandafter{\romannumeral 24} General Assembly
\cite{IAU:2001,Rickman:2001,Soffel:et:al:2003} of the IAU specifies a global
reference system, the Barycentric Celestial Reference System (BCRS),
and a physically adequate local geocentric reference system, the
Geocentric Celestial Reference System (GCRS), in the framework of the
post-Newtonian approximation of general relativity. Below it is argued
that a simplified version of the GCRS constructed for a massless
observer as a central body can be used as a physically adequate CoMRS.

The problem of constructing a physically adequate proper reference
system of a massive body (e.g. Earth) in the first post-Newtonian
approximation has been thoroughly discussed by several authors. In the
framework of general relativity two advanced formalisms are available.
One formalism is due to Brumberg and Kopeikin
\cite{Kopeikin:1988,Brumberg:Kopeikin:1989,Kopeikin:1990,Brumberg:1991,Klioner:Voinov:1993}
and another one is due to Damour, Soffel and Xu
\cite{Damour:Soffel:Xu:1991,Damour:Soffel:Xu:1992,Damour:Soffel:Xu:1993,Damour:Soffel:Xu:1994}.
For the gravitational $N$-body problem both formalisms introduce a
total of $N+1$ different coordinate systems: one set of global
coordinates $(t,x^i)$ and one set of local comoving coordinates
$(T, X^a)$ for each body. Note that in this context a body is just a
material subsystem at the boundaries of which the energy-momentum tensor vanishes.
In the local coordinates the metric tensor possesses the following two
properties:

\begin{itemize}

\item[{\bf A.}] The gravitational field of external bodies is
represented only in the form of a relativistic tidal potential which is
at least of second order in the local spatial coordinates and coincides
with the usual Newtonian tidal potential in the Newtonian limit.

\item[{\bf B.}] The internal gravitational field of the central body
coincides with the gravitational field of a corresponding isolated
source provided that the tidal influence of the external matter is
neglected.

\end{itemize}

These two requirements can be simultaneously satisfied in general
relativity as has been shown in the framework of the Brumberg-Kopeikin
and Damour-Soffel-Xu formalisms. It is clear that this fact is closely
related to the validity of the Strong Equivalence Principle in general
relativity. These two formalisms complement each other by elaborating
the theory from slightly different points of view. The formalisms
deliver (1) an elegant description of metric tensors in both the global
and local coordinates and the closed-form transformations between them,
(2) an improved description of the structure of the gravitational field
of each body by means of a set of its multipole moments which are
linked in an operational way to what can be observed in the local
gravitational environment of the body, (3) a description of the
influence of the external gravitational fields in the local reference
system by means of some suitably defined tidal moments, (4)
post-Newtonian translational and rotational equations of motion of the
$N$ bodies with full multipole structure, (5) physically adequate
equations of motion of a test particle in a local reference system, (6)
physically adequate relativistic models for many kinds of observations
(VLBI, high-accuracy positional observations, remote clock
synchronization, etc.). The IAU 2000 Resolution B1.3
\cite{Soffel:et:al:2003} that defines the metric tensors of both the
global BCRS and the local GCRS, is based on the results on these two
approaches.

In the framework of the PPN formalism with two parameters $\beta$ and
$\gamma$ the theory of physically adequate local reference systems was
developed in \cite{Klioner:Soffel:1998b,Klioner:Soffel:2000}. It is
clear that because of possible violation of the Strong Equivalence
Principle in some alternative theories of gravity it is, generally
speaking, impossible to construct a local reference system possessing
both properties {\bf A} and {\bf B}. In \cite{Klioner:Soffel:2000} it
has ben shown how to construct local reference systems which possess
either property {\bf A} or property {\bf B}. It has been also
demonstrated that for relativistic modeling of astronomical
observations one should normally prefer the local reference system with
property {\bf A}. The theory of local PPN reference systems with the
PPN parameters has been developed in
\cite{Klioner:Soffel:1998b,Klioner:Soffel:2000} with the same degree of
details as it was done in general relativity. For the limit of general
relativity $\beta=\gamma=1$ the formulas of \cite{Klioner:Soffel:2000}
coincide with those of the Brumberg-Kopeikin and Damour-Soffel-Xu
formalisms as well as with the formulas from the IAU 2000 Resolutions.

The problem of defining physically adequate local coordinates for a
massless body (test observer) is much more simple than the problem for
a massive body (e.g. for Earth). This problem has been discussed many
times in the literature. Let us mention, for example, the work of Ni
and Zimmermann \cite{Ni:Zimmermann:1978} where a local reference system
of an accelerated observer has been constructed explicitly up to the
terms of third order relative to the local spatial coordinates. As it
has been noted in \cite{Klioner:Voinov:1993}, the results of the
Brumberg-Kopeikin and Damour-Soffel-Xu formalisms can be directly
applied to define in an elegant way a physically adequate local
reference system of a massless body. Indeed, one should just consider
the limit where the gravitational potential of the central body
vanishes. It is clear that the same procedure can be applied also to
the local PPN reference system \cite{Klioner:Soffel:2000}. Exactly this
will be done below. The resulting reference system represents a natural
choice for the physically adequate CoMRS of a massless observer. This
reference system can be applied to model physical phenomena in the
immediate vicinity of the observer. Two examples will be given below:
the relation between the observed direction toward a light source and
the relevant BCRS parameters of the light ray, and the rotational
motion of the test observer (satellite).

Let us summarize the most important notations used throughout the
paper:

\begin{itemize}

\item $G$ is the Newtonian constant of gravitation;

\item $c$ is the velocity of light;

\item $\beta$ and $\gamma$ are the parameters of the parametrized post-Newtonian
(PPN) formalism which characterize possible deviation of the physical
reality from general relativity theory ($\beta=\gamma=1$ in general
relativity);

\item the lower case latin indices $i$, $j$, $k$, \dots take values
1, 2, 3;

\item the lower case Greek indices $\alpha$, $\beta$, \dots take values
0, 1, 2, 3;

\item repeated indices imply the Einsteinian summation irrespective of
their positions (e.g. $a^i\,b^i=a^1\,b^1+a^2\,b^2+a^3\,b^3$,
$a^\alpha\,b_\alpha=a^0\,b_0+a^1\,b_1+a^2\,b_2+a^3\,b_3$);

\item parentheses surrounding a group of indices denote symmetrization,
e.g., $A_{i(jk)}={1\over 2}\left(A_{ijk}+A_{ikj}\right)$;

\item brackets surrounding two indices denote antisymmetrization, e.g.,
$A_{i[jk]}={1\over 2}\left(A_{ijk}-A_{ikj}\right)$;

\item a comma before an index designates the partial derivative with
respect to the corresponding coordinates: $A_{,\mu}={\partial
A/\partial x^\mu}$, $A_{,i}={\partial A(t,\ve{x})/\partial x^i}$; for
partial derivatives with respect to the coordinate times $t$ and $\T$
we use special notations $A_{,t}={\partial A(t,\ve{x})/\partial t}$ and
$A_{,\T}={\partial A/\partial \T}$;

\item a dot over any quantity designates the total derivative with
respect to the coordinate time of the corresponding reference system:
e.g. $\dot A=\displaystyle{dA\over dt}$.

\end{itemize}

Sections~\ref{Section-PPN-BCRS}, \ref{Section-PPN-CoMRS} and
\ref{Section-PPN-CoMRS-BCRS-transformation} are devoted to the
definitions of the metric tensors of the BCRS and the CoMRS, and the
transformations between these two reference system, respectively. The
tetrad induced by the CoMRS coordinates at the CoMRS origin is
discussed in Section~\ref{Section-PPN-CoMRS-tetrad}.
Section~\ref{Section-aberration} elucidates the equivalence of several
possible ways to calculate the observable direction toward a light
source from the relevant coordinate quantities defined in the global
BCRS. The post-Newtonian equations of rotational motion of a satellite
relative to the CoMRS are briefly discussed in
Section~\ref{Section-rotational-motion}. In
Section~\ref{Section-attitude} it is argued that the kinematically
non-rotating CoMRS represents a natural choice of a reference system
where the attitude of the observer (e.g. of the GAIA satellite) is
modeled. A summary of the main results are given in
Section~\ref{Section-Conclusions}.

\section{The PPN metric tensor in the BCRS}
\label{Section-PPN-BCRS}

Let us consider an isolated system of $N$ gravitating bodies. It is
clear that the space-time is asymptotically flat and can be covered
with a single global coordinate system $x^\mu = (ct, x^i)$ where

\begin{equation}\label{BRS-limits}
\lim_{|\ve{x}|\rightarrow \infty \atop{t = {\rm const}}} g_{\mu\nu} =
\eta_{\mu\nu},
\end{equation}

\noindent
$g_{\mu\nu}$ being the metric tensor in the global coordinate system.
Here $\eta_{\mu\nu}={\rm diag}(-1,+1,+1,+1)$ is the Minkowski
metric tensor. In the framework of the PPN formalism
\cite{Will:1993} with two parameters $\beta$ and $\gamma$
the metric tensor $g_{\mu\nu}$ in the global
reference system can be written as
\cite{Klioner:Soffel:1998b,Klioner:Soffel:2000}

\begin{eqnarray}\label{BRS:metric}
g_{00}&=&-1+{2\over c^2}\,w(t,\ve{x})-{2\over c^4}
\,\beta\, w^2(t,\ve{x})+\OO5,
\nonumber \\
g_{0i}&=&-{2\,(1+\gamma)\over c^3}\,w^i(t,\ve{x})+\OO5,
\nonumber \\
g_{ij}&=&\delta_{ij}\left(1+{2\over c^2}\,\gamma\,
w(t,\ve{x})\right)+\OO4,
\end{eqnarray}

\noindent
where $\delta_{ij}$ is the Kronecker symbol. Here and below
3-dimensional coordinate quantities (``3-vectors'') referred to the
spatial axes of the corresponding reference system are set in boldface:
e.g. $\ve{x}=x^i$. A harmonic-like gauge for the global PPN metric
tensor is adopted here. Precisely speaking, the global metric tensor
satisfies the usual harmonic gauge ($g={\rm det}(g_{\mu\nu})$)

\begin{equation}\label{harmonic-gauge}
{\partial\over \partial x^\alpha}
\left( {(-g)}^{1/2} g^{\alpha\beta} \right)=0
\end{equation}

\noindent
in case of general relativity $\beta=\gamma=1$. This requires

\begin{equation}\label{w,t+wi,i=0}
w_{,t}+w^i_{,i}=\OO2.
\end{equation}

\noindent
In accordance with the standard PPN framework as described, e.g., in
\cite{Will:1993} the metric potentials $w$ and $w^i$ are assumed to
obey the equations

\begin{eqnarray}\label{UW,ii-U,tt}
w_{,ii}-{1\over c^2}\,w_{,tt}=-4\,\pi\,G\,\sigma+\OO4,
\\ \label{Uijj}
w^i_{,jj}=-4\,\pi\,G\,\sigma^i+\OO2,
\end{eqnarray}

\noindent
where

\begin{eqnarray}\label{sigma}
\sigma&=&{1\over c^2}\,\left(T^{00}+\gamma\,T^{kk}+
{1\over c^2}\,T^{00}\,(3\gamma-2\beta-1)\,w\right)+\OO4,
\\ \label{sigmai}
\sigma^i&=&{1\over c}\,T^{0i}+\OO2.
\end{eqnarray}

\noindent
Here, $T^{\mu\nu}$ are the components of the energy-momentum tensor in
the global reference system and $w$ in (\ref{sigma}) is needed  only to
Newtonian order where it coincides with the Newtonian potential.
Because of requirement (\ref{BRS-limits}) the solution of
(\ref{UW,ii-U,tt})--(\ref{Uijj}) can be written in the form

\begin{equation}\label{w-mu-solution}
w^\mu(t,\ve{x})= G \int {\sigma^\mu (t,\ve{x}') \over |\ve{x}-\ve{x}'|}d^3x'
+ {1\over 2 c^2}\,G\,{\partial^2 \over \partial t^2}\,
\int \sigma^\mu(t,\ve{x}') | \ve{x} - \ve{x}' | d^3x'+\OO4,
\end{equation}

\noindent
where $w^0=w$ and $\sigma^0=\sigma$. It is clear that this formulas for
$w$ and $w^i$ together with the Newtonian continuity equation

\begin{equation}\label{Newton-continuity}
\sigma_{,t}+\sigma^i_{,i}=\OO2
\end{equation}

\noindent
are compatible with the gauge condition (\ref{w,t+wi,i=0}). The metric
(\ref{BRS:metric})--(\ref{w-mu-solution}) is equivalent to the PPN
metric with coordinates $(t_{\rm pN},x^i_{\rm pN})$
in the standard post-Newtonian gauge
as used, e.g., in \cite{Will:1993}
up to a trivial gauge transformation

\begin{eqnarray}\label{tx-tx-PPN}
t_{\rm pN}&=&t-{1\over c^4}\,\chi_{,t}+\OO5,
\nonumber \\
x^i_{\rm pN}&=&x^i,
\end{eqnarray}

\noindent
where $\chi$ is the superpotential

\begin{equation}\label{chi}
\chi={1\over 2}\,G\,\int \sigma(t,\ve{x}')\,|\ve{x}-\ve{x}'|\,d^3x'+\OO2,
\end{equation}

\noindent
so that

\begin{equation}\label{chi,ii=w}
\chi_{,ii}=w+\OO2.
\end{equation}

For $\beta=\gamma=1$ the definition (\ref{BRS:metric}) and
(\ref{sigma})--(\ref{w-mu-solution}) coincides with the definition of
the BCRS \cite{IAU:2001,Soffel:et:al:2003} given within general
relativity.

 \section{The PPN metric tensor in the CoMRS}
\label{Section-PPN-CoMRS}

The CoMRS is a physically adequate reference system of an observer
$(\T,\X^a)$ the mass of which is so small that its influence on the
background space-time can be neglected. The metric tensor in the CoMRS
can be derived from the metric tensor of the GCRS by setting the
gravitational potential of the central body to zero. Below we modify in
this way the PPN version of the GCRS as constructed in
\cite{Klioner:Soffel:2000}. Again by setting $\gamma=\beta=1$ in the
formulas below one can restore the formulas which could be derived
directly from the GCRS adopted by the IAU
\cite{IAU:2001,Rickman:2001,Soffel:et:al:2003}. The metric tensor in
the CoMRS reads

\begin{eqnarray}\label{GRS:metric}
\G_{00}&=&-1+{2\over c^2}\,\W(\T,\ve{\X})-{2\over c^4}
\,\beta\,\W^2(\T,\ve{\X})+\OO5,
\nonumber \\
\G_{0a}&=&-{2(1+\gamma)\over c^3}\,\W^a(\T,\ve{\X})+\OO5,
\nonumber \\
\G_{ab}&=&\delta_{ab}\left(1+{2\over c^2}\,\gamma\,
\W(\T,\ve{\X})\right)+\OO4,
\end{eqnarray}

\noindent
where

\begin{eqnarray}\label{GRS:split}\label{W-split}
\W(\T,\ve{\X})&=&\Q_a(\T)\,\X^a+\W_{\rm T}(\T,\ve{\X}),
\\ \label{Wi-split}
\W^a(\T,\ve{\X})&=&{1\over 2}\,\varepsilon_{abc}\,\C_b(\T)\,\X^c
+\W^a_{\rm T}(\T,\ve{\X}),
\end{eqnarray}

\noindent
where  $\varepsilon_{ijk}=(i-j)(j-k)(k-i)/2$ is the fully antisymmetric
Levi-Civita symbol. Potentials $\W_{\rm T}(\T,\ve{\X})$ and $\W^a_{\rm
T}(\T,\ve{\X})$ are the external tidal gravitational potentials (both
are ${\cal O}(|\ve{\X}|^2)$) which describe the manifestation of the
external gravitational field in the CoMRS. The terms $\Q_a(\T)\,\X^a$
and ${1\over 2}\,\varepsilon_{abc}\,\C_b(\T)\,\X^c$, $\Q_a$ and $\C_a$
being arbitrary functions of time $\T$, are linear relative to
$|\ve{\X}|$ and describe the translational and rotation motion of the
CoMRS. The $\Q_a$ is the acceleration of
the momentarily co-moving locally inertial reference system
relative to
the CoMRS origin.
In other words, an accelerometer placed at the CoMRS origin
measures $-\Q_a$ (see, Section VIII of \cite{Klioner:Soffel:2000} where
the equations of motion of a test particle relative to the local PPN
coordinates were derived). If the observer (satellite) is a drag-free
satellite one can set $\Q_a=0$. If the observer (satellite) is equipped
with some kind of thrusters, $\Q_a(\T)$ can be used to describe their
influence. Non-gravitational forces can be also described by choosing
some special model for $\Q_a(\T)$. In the following we consider $\Q_a$
as arbitrary function. The $\C_a$ defines the rotational motion of the
spatial axes of the CoMRS relative to the momentarily co-moving
Fermi-Walker transported locally
inertial reference system. Clearly, the equations of test particles
relative to the CoMRS with $\C_a\neq0$ contain Coriolis forces.
Possible choices of $C_a$ and its relation to the rotational matrix
$\R^a_{\ i}$ in the transformation between the CoMRS spatial
coordinates and the spatial coordinates of the BCRS will be discussed
below.


Here the harmonic gauge conditions are again assumed to be valid in the
$\beta=\gamma=1$ limit. This implies an equation similar to
(\ref{w,t+wi,i=0}) for potentials $\W$ and $\W^a$ that in turn gives

\begin{equation}\label{WT,u+Qi,uXi+WTi,i=0}
\dot Q_a\,\X^a+\W_{{\rm T},\T}+\W^a_{{\rm T},a}=\OO2.
\end{equation}

\noindent
Now from the results of \cite{Klioner:Soffel:1998b,Klioner:Soffel:2000}
one gets

\begin{eqnarray}\label{W_T}
\W_{\rm T}(\T,\ve{\X})=&&
w(t,\ve{x})-w(\ve{x}_\o)
-w_{,j}(\ve{x}_\o)\,r_\o^j
\nonumber \\
&&
+{1\over c^2}\biggl(
-2(1+\gamma)\, v_\o^i \,
\left(w^i(t,\ve{x})-w^i(\ve{x}_\o)
-w^i_{,j}(\ve{x}_\o)\,r_\o^j \right)
\nonumber \\
&&
\phantom{+{1\over c^2}\biggl(}
+(1+\gamma)\, v_\o^2\,
\left(w(t,\ve{x})-w(\ve{x}_\o)-w_{,j}(\ve{x}_\o)\,r_\o^j\right)
\nonumber \\
&&
\phantom{+{1\over c^2}\biggl(}
+(1+\gamma)\,\dot{w}^{i,j}(\ve{x}_\o)\,r_\o^i
\,r_\o^j
+{1\over 2}\,\gamma\,\ddot{w}(\ve{x}_\o)\,r_\o^2
+\left({1\over 2}-\beta-\gamma\right)\,{(a_\o^i r_\o^i)}^2
\nonumber \\
&&
\phantom{+{1\over c^2}\biggl(}
+(1-2\beta-2\gamma)\,\Q_a\,\R^a_i\,r_\o^i\,a_\o^j\,r_\o^j
-\gamma\,v_\o^i\,r_\o^i\,
\dot{w}_{,j}(\ve{x}_\o)\,r_\o^j
\nonumber \\
&&
\phantom{+{1\over c^2}\biggl(}
+{1\over 2}\gamma\,r_\o^2\,\Q_a\,\R^a_{\ i}\,a_\o^i
+{1\over 10}\,(\gamma-2)\,r_\o^2\,\ddot a_\o^i\,r_\o^i
\nonumber \\
&&
\phantom{+{1\over c^2}\biggl(}
+2(1-\beta)\left[w(\ve{x}_\o)
+a_\o^i\,r_\o^i\right]\,
\left(w(t,\ve{x})-w(\ve{x}_\o)
-w_{,j}(\ve{x}_\o)r_\o^j\right)\,
\biggr)
\nonumber \\
&&
+\OO4,
\end{eqnarray}

\begin{eqnarray}\label{W_T^i:simplest}
\W_{\rm T}^a(\T,\ve{\X})=&&\R^a_{\ i}\,\biggl\{w^i(t,\ve{x})-
w^i(\ve{x}_\o)
-w^i_{,j}(\ve{x}_\o)\,r_\o^j
-v_\o^i\,
\left(w(t,\ve{x})-w(\ve{x}_\o)-w_{,j}(\ve{x}_\o)r_\o^j\right)
\nonumber \\
&&
\phantom{\R^a_{\ i}\,\biggl\{}
+{2\gamma+1\over 5\,(1+\gamma)}\,r_\o^i\,\dot a_\o^j\,r_\o^j
-{3\gamma-1\over 10\,(1+\gamma)}\,{\dot{a}_\o}^i\,\ve{r}_\o^2
\biggl\}
+\OO2.
\end{eqnarray}

\noindent
Here $r_\o^i=x^i-x_\o^i(t)$, $x_\o^i(t)$ are the coordinates of the
origin of the local reference system relative to the global one, and
$v_\o^i=dx_\o^i/dt$ and $a_\o^i=d^2x_\o^i/dt^2$ are its velocity and
acceleration, respectively. For any function of $A(t,\ve{x})$ we use
the shorthand notation $A(\ve{x}_\o)=A(t,\ve{x}_\o(t))$.

\section{Transformation from the BCRS to the CoMRS}
\label{Section-PPN-CoMRS-BCRS-transformation}

The coordinate transformations between the BCRS and the CoMRS read
\cite{Klioner:Soffel:1998b,Klioner:Soffel:2000}

\begin{eqnarray}
\label{trans:time}
&&\T=t-{1\over c^2}\left(\A+v_\o^i\,r_\o^i\right)+
{1\over c^4}\left(\B+\B^i\,r_\o^i+ \B^{ij}\, r_\o^i\, r_\o^j
+\C(t,\ve{x})
\right)+{\cal O}(c^{-5}),
\\ \label{trans:space}
&&\X^a=\R^a_{\,j}\left(r_\o^j+{1\over c^2}
\left({1\over 2}v_\o^j\,v_\o^k\,r_\o^k+\D^{jk}\,r_\o^k
+\D^{jkl}\,r_\o^k\,r_\o^l\right)\right)
+\OO4,
\end{eqnarray}

\begin{equation}\label{A}
{\dot \A}(t)={1\over 2}v_\o^2+w(\ve{x}_\o),
\end{equation}

\begin{equation}\label{B}
{\dot \B}(t)=-{1\over 8}\,v_\o^4
+2(\gamma+1)\,v_\o^i\,w^i(\ve{x}_\o)
-\left(\gamma+{1\over 2}\right)\,v_\o^2\,w(\ve{x}_\o)
+\left(\beta-{1\over 2}\right)\,
w^2(\ve{x}_\o),
\end{equation}

\begin{equation}\label{B^i}
\B^i(t)=-{1\over 2}\,v_\o^2\,v^i_\o+
        2(1+\gamma)\,w^i(t,\ve{x}_\o)-
        (2\gamma+1)\,v_\o^i\,w(\ve{x}_\o),
\end{equation}

\begin{equation}\label{B^ij}
\B^{ij}(t)= -v_\o^{(i}\,\R^a_{\,j)}\,\Q^{a}+
(1+\gamma)w^{(i,j)}(\ve{x}_\o)
-\gamma\,v_\o^{(i}\,w^{,j)}(\ve{x}_\o)+{1\over 2}\,\gamma
\,\delta^{ij}\,\dot{w}(\ve{x}_\o),
\end{equation}

\begin{equation}\label{C_1}
\C(t,\ve{x})={1\over 10}\,(\gamma-2)\,r_\o^2\,({\dot{a}_\o}^i\,r_\o^i),
\end{equation}

\begin{equation}\label{D^{ij}}
\D^{ij}(t)=\delta^{ij}\,\gamma\,w(\ve{x}_\o),
\end{equation}

\begin{equation}\label{D^{ijk}}
\D^{ijk}(t)={1\over 2}\,\gamma\,\left(\delta^{ij}a_\o^k+
\delta^{ik}a_\o^j-\delta^{jk}a_\o^i\right).
\end{equation}

Besides that, the rotational matrix $\R^a_i$ from the transformation of
the spatial coordinates (\ref{trans:space}) is related to $\C_a$ from
(\ref{Wi-split}) as

\begin{eqnarray} \label{Rij-Ca}
c^2\,\R^a_{\ i}\,\dot{\R}^a_{\ j}&=&-(1+\gamma)\,\varepsilon_{ijk}\,
\R^a_{\ k}\,\C_a
\nonumber \\
&&
+(1+2\gamma)\,v_\o^{[i}\,w_{,j]}(\ve{x}_\o)
-2(1+\gamma)\,w^{[i,j]}(\ve{x}_\o)
+v_\o^{[i}\,\R^a_{\ j]}\,\Q_a+\OO2.
\end{eqnarray}

\noindent
The dynamically non-rotating version of the CoMRS characterizes by
$\C_a=0$ (i.e. no Coriolis forces in the equations of motion of test
particles) and, as it follows from (\ref{Rij-Ca}), has specific spatial
rotation relative to the BCRS consisting of geodetic (de Sitter),
Lense-Thirring and Thomas precessions. The GCRS of the IAU is defined
to be kinematically non-rotating, that is, it has no rotation of
spatial axes relative to the BCRS (i.e. $\R^a_i=\delta^a_i$ in this
case). Although Coriolis forces appear in the equations of motion
relative to a kinematically non-rotating reference system, this choice
is especially advantageous for modeling of astronomical observations,
since no additional orientation-related re-calculations (e.g., of
planetary ephemeris data) are necessary. The same arguments apply to
the CoMRS: the kinematically non-rotating CoMRS is the most convenient
choice of the orientation of the local coordinates. For the
kinematically non-rotating CoMRS $\R^a_i=\delta^a_i$ and $\C_a$ has
some specific non-zero value defined by (\ref{Rij-Ca}). Below we retain
$\R^a_i$ in the formulas, but it should be chosen to be identity matrix
$\delta^a_i$.

Matching of the CoMRS and BCRS metric tensors allows one to derive also
the equations of motion of the CoMRS origin as well. The BCRS
acceleration of the CoMRS origin (that is, the acceleration
of the observer's center of mass) reads

\begin{eqnarray}\label{aei}
&&a_\o^i=w_{,i}(\ve{x}_\o)+\Delta a_\o^i
\nonumber \\
&&
\phantom{a_\o^i=}
+{1\over c^2}\biggl(
2(1+\gamma)\,\dot{w}^i(\ve{x}_\o)
+\left(\gamma v_E^2
-2(\gamma+\beta)\,w(\ve{x}_\o)\right)
w_{,i}(\ve{x}_\o)
-(2\gamma+1)\,v_\o^i\,\dot{w}(\ve{x}_\o)
\nonumber \\
&&
\phantom{a_\o^i(t^*)=+{1\over c^2}\biggl(}
-2(1+\gamma)\,v_\o^j\,w^j_{,i}(\ve{x}_\o)
-v_\o^i\,v_\o^j\,w_{,j}(\ve{x}_\o)
\biggr)
+{\cal O}(c^{-4}),
\end{eqnarray}

\noindent
where

\begin{eqnarray}\label{delta-aei}
\Delta a_\o^i&=&-\R^a_j\,\Q_a
\left(\delta^{ij}-{1\over c^2}\left(
v_E^2\,\delta^{ij}
+(2+\gamma)\,w(\ve{x}_\o)\,\delta^{ij}
+{1\over 2} v_\o^i\,v_\o^j
\right)\right)
\end{eqnarray}

\noindent
is the BCRS coordinate acceleration of the CoMRS origin relative to the
co-moving locally inertial reference system. Clearly, $\Delta a_\o^i$
is proportional to $\Q_a$ and comes just from the re-calculation of the
CoMRS-defined $\Q_a$ into the BCRS (see, the discussion of Eq. (4.29) of
\cite{Klioner:Soffel:2000}). If the acceleration $\Delta a_\o^i$ is
neglected, Eq. (\ref{aei}) coincides with the equation of time-like
geodetic in the metric (\ref{BRS:metric}). If the gravitational fields of
all $N$ bodies can be described only by their masses (no further
multipole moments of the gravitational field in the corresponding local
reference system of each body), Eq. (\ref{aei}) produces Eq. (3) of
\cite{Klioner:2003}.

From (\ref{trans:time}) with (\ref{A}) and (\ref{B}) follows that the
CoMRS coordinate time $\T$ at the CoMRS origin $\X^a=0$ (this is
equivalent to $r_\o^i=0$) coincides with the proper time $\tau$ of the
test observer.

\section{Coordinate-induced tetrad for the origin of the CoMRS}
\label{Section-PPN-CoMRS-tetrad}

Let us construct a tetrad (e.g., \cite{Landau:Lifshitz:1971})
co-moving with the observer. Let us first introduce four vectors ${\rm
e}_{(\alpha)}^\mu$ attached to a point on the worldline of the observer.
Here index $\alpha$ is the tetrad index which runs
from 0 to 3 and numerates the vectors. Index $\mu$ is a normal tensor
index which can be lowered and raised by contracting with the metric
tensor

\begin{eqnarray}\label{tensor-indices-tetrad}
{\rm e}_{(\alpha)\,\mu}&=&g_{\mu\nu}\,{\rm e}_{(\alpha)}^\nu,
\nonumber\\
{\rm e}_{(\alpha)}^\mu&=&g^{\mu\nu}\,{\rm e}_{(\alpha)\,\nu}.
\end{eqnarray}

\noindent
The four vectors are required to have the property

\begin{equation}\label{tetrad-condition-1}
{\rm e}_{(\alpha)}^\mu\,{\rm e}_{(\beta)\,\mu}=\eta_{\alpha\beta}.
\end{equation}

\noindent
This equation implies that the vectors are orthogonal to each other,
that vector ${\rm e}_{(0)}^\mu$ is unit and time-like, and that ${\rm
e}_{(i)}^\mu$ are unit and space-like. Four additional vectors ${\rm
e}^{(\alpha)\,\mu}$ (with tetrad index written as superscript) are
then defined by

\begin{equation}\label{tetrad-condition-2}
{\rm e}^{(\alpha)\,\mu}\,{\rm e}_{(\beta)\,\mu}=\delta^\alpha_\beta,
\end{equation}

\noindent
$\delta^\alpha_\beta={\rm diag}(1,1,1,1)$ is the 4-dimensional Kronecker
symbol. From (\ref{tetrad-condition-1}) and (\ref{tetrad-condition-2})
it is easy to show that the conversion between vectors
${\rm e}_{(\alpha)}^\mu$ and ${\rm e}^{(\alpha)\,\mu}$ can be
performed simply by contracting with the Minkowski metric

\begin{eqnarray}\label{tetrad-indices-tetrad}
{\rm e}_{(\alpha)}^\mu&=&\eta_{\alpha\beta}\,{\rm e}^{(\beta)\,\mu},
\nonumber\\
{\rm e}^{(\alpha)\,\mu}&=&\eta^{\alpha\beta}\,{\rm e}_{(\beta )}^\mu,
\end{eqnarray}

\noindent
where $\eta^{\alpha\beta}={\rm diag}(-1,1,1,1)$ is the matrix inverse
to $\eta_{\alpha\beta}$. With the help of these vectors one can
represent the metric tensor at the considered point of space-time as

\begin{equation}\label{g-mu-nu-tetrad}
g_{\mu\nu}={\rm e}^{(\alpha)}_\mu\,{\rm e}_{(\alpha)\,\nu}=
\eta_{\alpha\beta}\,{\rm e}^{(\alpha)}_\mu\,{\rm e}^{(\beta)}_\nu,
\end{equation}

\noindent
so that

\begin{equation}\label{tetrad-locally-cartesian}
ds^2=\eta_{\alpha\beta}\,dx^{(\alpha)}\,dx^{(\beta)},
\end{equation}

\noindent
with

\begin{equation}\label{dx-tetrad}
dx^{(\alpha)}={\rm e}^{(\alpha)}_\mu\,dx^\mu.
\end{equation}

\noindent
Eq. (\ref{tetrad-locally-cartesian}) shows that $dx^{(\alpha)}$ can be
interpreted as observable infinitely small time intervals and
distances in the infinitesimal neighborhood of the considered space-time point.
For a tetrad co-moving with the observer the unit time-like vector
${\rm e}_{(0)}^\mu$ can be chosen to coincide with the 4-velocity of
the observer, so that the projection $dx^{(0)}={\rm
e}^{(0)}_\mu\,dx^\mu$ coincides with the interval of the proper time of
the observer $dx^{(0)}=d\tau$ between the events with coordinates
$x^\mu$ and $x^\mu+dx^\mu$ both lying on the worldline of the observer. This means

\begin{equation}\label{monad}
{\rm e}^{(0)}_\mu=-g_{\mu\nu}\,{dx^\nu_\o\over c\,d\tau}.
\end{equation}

\noindent
The vectors ${\rm e}^{(i)\,\mu}$ are then constrained by
(\ref{g-mu-nu-tetrad}) up to arbitrary spatial rotation. This means
that if ${\rm e}^{(i)}_\mu$ is a solution
of (\ref{g-mu-nu-tetrad})
then

\begin{equation}
\label{tetrad-rotation}
\overline{\rm e}^{(k)}_\mu=P^k_j\,{\rm e}^{(j)}_\mu
\end{equation}

\noindent
with arbitrary orthogonal matrix $P^k_j$ is also a solution.

Considering the BCRS metric tensor (\ref{BRS:metric}) one gets
from (\ref{monad})

\begin{eqnarray}\label{monad-bcrs}
{\rm e}^{(0)}_0&=&
1
+{1\over c^2}\,\left({1\over 2}\,v_\o^2-w(\ve{x}_\o)\right)
\nonumber\\
&&\phantom{1}
+{1\over c^4}\,\left({3\over 8}\,v_\o^4
                     +\left(\gamma+{1\over 2}\right)\,v_\o^2\,w(\ve{x}_\o)
                     +\left(\beta-{1\over 2}\right)\,w^2(\ve{x}_\o)\right)
+\OO5,
\\
\label{monad-bcrs-i}
{\rm e}^{(0)}_i&=&
-{1\over c}\,v_\o^i
+{1\over c^3}\,\left(
-{1\over 2}\,v^2_\o\,v^i_\o
-(2\,\gamma+1)\,w(\ve{x}_\o)\,v^i_\o
+2\,(1+\gamma)\,w^i(\ve{x}_\o)
\right)
+\OO5.
\end{eqnarray}

\noindent
The following partial solution for ${\rm e}^{(a)}_\mu$ is then
possible:

\begin{eqnarray}\label{triad-bcrs0}
{\rm e}^{(a)}_0&=&-{1\over c}\,v^a_\o\,\left(1+{1\over c^2}\left({1\over 2}\,v_\o^2+
\gamma\,w(\ve{x}_\o)\right)\right)+\OO5,
\\
\label{triad-bcrsa}
{\rm e}^{(a)}_i&=&\delta^{ai}+{1\over c^2}\,\left({1\over 2}\,v^a_\o\,v^i_\o+
\gamma\,w(\ve{x}_\o)\,\delta^{ai}\right)+\OO4.
\end{eqnarray}

This solution for ${\rm e}^{(a)}_i$ shows that we have chosen the
space-like vectors of the tetrad is such a way that they show no
spatial rotation relative to the spatial axes of the BCRS. It is easy
to see from (\ref{trans:time})--(\ref{D^{ijk}}) that this tetrad is the
coordinate basis of the kinematically non-rotating CoMRS
(i.e. of the CoMRS with $\R^a_i=\delta^a_i$)
on the worldline of its origin

\begin{eqnarray}\label{tetrad-CoMRS}
{\rm e}^{(\alpha)}_\mu&=&\left.
{\partial \X^\alpha\over \partial x^\mu}
\right|_{\X^i=0}.
\end{eqnarray}

\noindent
Therefore, the CoMRS implies also adopting a particular tetrad co-moving
with the observer. Tetrad (\ref{monad-bcrs})--(\ref{triad-bcrsa}) is
induced by the CoMRS coordinates at the origin of the CoMRS in the sense
of Section 3.4.2 of \cite{Soffel:1988}. This tetrad can be used to
model certain kind of observables (e.g., proper directions as
directions relative to the tetrad
(\ref{monad-bcrs})--(\ref{triad-bcrsa})). However, the CoMRS is more
than just a tetrad and allows one to use all the power of the theory of
local reference systems as mentioned in Section
\ref{Section-Introduction}.

If one adopts a dynamically non-rotating CoMRS with $\C_a\equiv0$ and
$\R^a_i$ defined by (\ref{Rij-Ca}), the corresponding tetrad
(\ref{tetrad-CoMRS}) will be Fermi-Walker transported, so that the
Fermi rotation coefficients of that tetrad vanish. This is, however, an
unnecessary complication for space astrometry, where tetrad
(\ref{monad-bcrs})--(\ref{triad-bcrsa}) and the kinematically
non-rotating CoMRS are more convenient.

The tetrad (\ref{monad-bcrs})--(\ref{triad-bcrsa}) is written for arbitrary
velocity of the observer $v^i_\o$. The tetrads $\widetilde{\rm
e}^{(\alpha)}_\mu$ and ${\rm e}^{(\alpha)}_\mu$ defined by
(\ref{monad-bcrs})--(\ref{triad-bcrsa}) with two different velocities
$\widetilde{v}^i_\o$ and $v^i_\o$, respectively, are related to each
other by a Lorentz transformation $\Lambda^\alpha_\beta$ plus
additional spatial rotation $\P^i_{\ j}$ of space-like vectors:

\begin{eqnarray}\label{tetrad-Lorentz}
{\rm e}^{(0)}_\mu&=&\Lambda^0_\alpha\,\widetilde{\rm e}^{(\alpha)}_\mu,
\\
{\rm e}^{(i)}_\mu&=&\P^i_{\ j}\,\Lambda^j_\alpha\,\widetilde{\rm e}^{(\alpha)}_\mu,
\end{eqnarray}

\noindent
where

\begin{eqnarray}\label{Lorentz-Lambdas}
\Lambda^0_0&=&\Gamma,
\nonumber\\
\Lambda^0_a&=&-{1\over c}\,\nu^a\,\Gamma,
\nonumber\\
\Lambda^i_0&=&-{1\over c}\,\nu^i\,\Gamma,
\nonumber\\
\Lambda^i_a&=&\delta^{ia}+{1\over c^2}\,{\Gamma^2\over 1+\Gamma}\,\nu^i\,\nu^a,
\nonumber\\
\Gamma&=&\left(1-{1\over c^2}\,\nu^k\,\nu^k\,\right)^{-1/2},
\end{eqnarray}

\noindent
and

\begin{eqnarray}\label{additional-P}
\P^i_{\ j}&=&\delta^{ij}+{1\over 2\,c^2}\,
\left(\widetilde{v}_\o^i\,v_\o^j-\widetilde{v}_\o^j\,v_\o^i\right)+\OO4.
\end{eqnarray}

\noindent
Here, the parameter $\nu^i$ of the Lorentz transformation is the BCRS
velocity $v^i_\o$ of the second observer as seen by the first observer
having the BCRS velocity $\widetilde{v}^i_\o$ which can be calculated
as

\begin{eqnarray}\label{Lorentz-nu}
\nu^i&=&
c\,{d\widetilde{x}^{(i)}\over d\widetilde{x}^{(0)}}=
c\,{\widetilde{\rm e}^{(i)}_\mu\,dx^\mu\over
\widetilde{\rm e}^{(0)}_\mu\,dx^\mu}=c\,
{\widetilde{\rm e}^{(i)}_0+\widetilde{\rm e}^{(i)}_j\,v^j_\o/c\over
\widetilde{\rm e}^{(0)}_0+\widetilde{\rm e}^{(0)}_k\,v^k_\o/c}
\nonumber\\
&=&v_\o^i-\widetilde{v}_\o^i
+{1\over c^2}\,\left(
(v_\o^j\,\widetilde{v}_\o^j)\,(v_\o^i-\widetilde{v}_\o^i)
-{1\over 2}\,(\widetilde{v}_\o^j\,\widetilde{v}_\o^j)\,v_\o^i
+{1\over 2}\,(v_\o^j\,\widetilde{v}_\o^j)\,\widetilde{v}_\o^i
\right.
\nonumber\\
&&
\left.
\phantom{v_\o^i-\widetilde{v}_\o^i+{1\over c^2}\,\Biggl(}
+(1+\gamma)\,w(\ve{x}_\o)(v_\o^i-\widetilde{v}_\o^i)
\right)+\OO4.
\end{eqnarray}

\noindent
In the limit of special relativity Eq. (\ref{Lorentz-nu})
coincides with the special-relativistic velocity composition law.
On the other hand, if the BCRS velocity of the first observer
vanishes ($\widetilde{v}^i_\o=0$), Eq. (\ref{Lorentz-nu})
reproduces Eq. (12) of \cite{Klioner:2003}.

The reason for the appearance of the additional spatial rotation
$\P^i_{\ j}$ is the well-known fact that two subsequent Lorentz
transformations without spatial rotation are equivalent to the one
Lorentz transformation with spatial rotation. This additional spatial
rotation and its consequences for kinematically non-rotating
astronomical reference systems are discussed in
\cite{Klioner:1993,Klioner:Soffel:1998a}. The matrix
(\ref{additional-P}) is equal to identity matrix $\delta^i_{\ j}$ if
$\widetilde{v}_\o^i=0$, so that if $\widetilde{v}_\o^i=0$ the
transformation between $\widetilde{\rm e}^{(\alpha)}_\mu$ and ${\rm
e}^{(\alpha)}_\mu$ is a pure Lorentz transformation with parameter
$\nu^i=v_\o^i\,\left(1+c^{-2}\,(1+\gamma)\,w(\ve{x}_\o)\right)+\OO4$.
Using the Lorentz transformation in its closed form allows one to get
the expressions for the tetrad in the first post-Minkowskian
approximation, that is, retaining terms of first order in $G$ and of
all orders in $|v^i_\o|/c$.

\section{Observed direction of the light propagation}
\label{Section-aberration}

Let us compute explicitly the relation between the unit coordinate
direction of light propagation $n^i$ in the BCRS and the observed
direction to the light source relative to the tetrad
(\ref{monad-bcrs})--(\ref{triad-bcrsa}). Let $x^\mu_p(t)$ be the
coordinates of the photon parametrized by the coordinate time $t$.
It is clear that the observed direction should be defined with respect
to the tetrad vectors ${\rm e}^{(\alpha)}_\mu$ as

\begin{equation}\label{s-(a)}
s^{(a)}=-{dx^{(a)}_p\over dx^{(0)}_p}=
- {{\rm e}^{(a)}_\mu\,dx^\mu_p\over
{\rm e}^{(0)}_\mu\,dx^\mu_p}
=
- {{\rm e}^{(a)}_0+{\rm e}^{(a)}_i\,p^i\over
{\rm e}^{(0)}_0+{\rm e}^{(0)}_j\,p^j},
\end{equation}

\noindent
where $p^i={1\over c}\,{dx^i_p\over dt}$ is the coordinate light
velocity at the point of observation $x^i=x^i_\o$. The differentials of
$x^\mu_p$ are calculated here along the light ray $x^\mu_p(t)$ at the
point of observation. However, because of (\ref{tetrad-CoMRS}) the
direction $s^{(a)}$ coincides with the CoMRS velocity of the light
propagation at the origin of the CoMRS (that is with the velocity $p^i$
transformed into the CoMRS using the coordinate transformation
(\ref{trans:time})--(\ref{trans:space})):

\begin{equation}\label{s-a}
s^a=-{d\X^{a}_p\over d\X^{0}_p}=
- {{\partial \X^a\over \partial x^\mu}\,dx^\mu_p\over
   {\partial \X^0\over \partial x^\mu}\,dx^\mu_p}
=
- {{\partial \X^a\over \partial x^0}+{\partial \X^a\over \partial x^i}\,p^i\over
   {\partial \X^0\over \partial x^0}+{\partial \X^0\over \partial x^j}\,p^j}
=s^{(a)}.
\end{equation}

\noindent
Here $\X^\alpha_p(\T)$ is the worldline $x^\mu_p(t)$ of the light ray
expressed in the CoMRS coordinates. All partial derivatives ${\partial
\X^\alpha\over \partial x^\mu}$ as well as the differentials
$d\X^\alpha_p$ are calculated at the point of observation. Let us also
note that (\ref{s-(a)}) and (\ref{s-a}) have one more interpretation.
Let us consider two 4-vectors $a^\mu$ and $b^\mu$ and an observer with
4-velocity $u^\mu={dx^\mu_\o\over c\,d\tau}$. Let us also assume that
both vectors $a^\mu$ and $b^\mu$ are not equal to $A\,u^\mu$, $A$ being
a constant. It is well known (e.g., \cite{Soffel:1988,Will:1993}) that
by projecting each of these vectors into the observer's rest space and
calculating the normalized scalar product of the projected vectors with
respect to the metric $g_{\mu\nu}$ one gets the cosine of the angle
$\theta$ between these two vectors as measured by the observer:

\begin{eqnarray}
\overline{a}^\mu&=&P_{\mu\nu}\,a^\nu,
\\
\overline{b}^\mu&=&P_{\mu\nu}\,b^\nu,
\\
\cos\theta&=&{\overline{a}_\mu\,\overline{b}^\mu\over
              \left(\overline{a}_\alpha\,\overline{a}^\alpha\right)^{1/2}\,
              \left(\overline{b}_\beta\,\overline{b}^\beta\right)^{1/2}},
\end{eqnarray}

\noindent
where $P_{\mu\nu}=g_{\mu\nu}+u_\mu\,u_\nu$ is the projection operator
into the satellite's rest space. Using (\ref{monad}) and
(\ref{g-mu-nu-tetrad}) one can write
$P_{\mu\nu}=g_{\mu\nu}+e^{(0)}_\mu\,e^{(0)}_\nu
=\delta_{ab}\,e^{(a)}_\mu\,e^{(b)}_\nu$. Therefore, the cosine of the
observed angle $\theta_a$ between the incoming light ray with the wave
vector $k^\mu={dx^\mu_p\over c\,dt}=(1,p^i)$ and a space-like
vector of the triad ${\rm e}^{(a)}_\mu$ can be calculated as

\begin{equation}
\label{theta-a}
\cos\theta_a=
{
{\rm e}^{(a)}_\mu\,k^\mu
\over
{\rm e}^{(0)}_\nu\,k^\nu
}.
\end{equation}

\noindent
Note that $P^{\alpha}_{\ \beta}\,{\rm e}^{(a)\,\beta}={\rm
e}^{(a)\,\alpha}$ and ${\rm e}^{(a)\,\alpha}\,{\rm
e}^{(b)}_\alpha=\delta^{ab}$ and, therefore, vectors ${\rm
e}^{(a)}_\alpha$ already lie in the observer's rest space and are
normalized to unity. In (\ref{theta-a}) we used also that according to
(\ref{tetrad-condition-1}) and (\ref{monad}) $u_\alpha\,{\rm
e}^{(a)\,\alpha}=0$ for any $a=1,2,3$. This technique to compute the
cosines of the observed angles has been used in a slightly different
form e.g. in \cite{Bini:deFelice:2003,Bini:Crosta:deFelice:2003}. It
is, however, clear that this technique is equivalent to the two
above-mentioned ways to derive $s^{(a)}$ and the components of
$s^{(a)}$ can be easily related to $\cos\theta_a$. Indeed, one has

\begin{eqnarray}
\label{theta-a-s-(a)}
\cos\theta_a&=&
{{\rm e}^{(a)}_\mu\,k^\mu\over {\rm e}^{(0)}_\nu\,k^\nu}
={{\rm e}^{(a)}_\mu\,dx^\mu_p\over {\rm e}^{(0)}_\nu\,dx^\nu_p}=-s^{(a)}.
\end{eqnarray}

\noindent
The difference in the sign between $\cos\theta_a$ and $s^{(a)}$ reflects
the fact that $s^{(a)}$ is the direction toward the source while
$\cos\theta_a$ characterizes the opposite direction of light propagation.

Now one can substitute (\ref{monad-bcrs})--(\ref{triad-bcrsa}) into
(\ref{s-(a)}) or (\ref{s-a}) and expand the denominator into powers of
$c^{-1}$ to get the explicit relation between $s^{(a)}$ and $p^i$. The
absolute value of the coordinate light velocity can be calculated from
the fact that the light follows a null geodetic which means that in a
reference system with metric tensor $g_{\alpha\beta}$ vector $p^i$
satisfies the equation

\begin{equation}\label{null-geodetic}
g_{\mu\nu}\,k^\mu\,k^\nu=g_{00} + 2 g_{0i}\,p^i + g_{ij}\,p^i\,p^j=0.
\end{equation}

\noindent
Substituting the BCRS metric (\ref{BRS:metric}) into (\ref{null-geodetic})
one gets

\begin{equation}\label{abs-p}
|\ve{p}|=1
-{1\over c^2}\,(1+\gamma)\,w(\ve{x}_\o)
+{1\over c^3}\,2\,(1+\gamma)\,n^i\,w^i(\ve{x}_\o)+\OO4,
\end{equation}

\noindent
where $|\ve{p}|=(p^1\,p^1+p^2\,p^2+p^3\,p^3)^{1/2}$ is the Euclidean norm
of $p^i$.
Combining (\ref{s-(a)})--(\ref{s-a})
 with (\ref{monad-bcrs})--(\ref{triad-bcrsa})
and (\ref{abs-p}) one gets

\begin{eqnarray}\label{aberration}
\ve{s}=-\ve{n}&+&\,{1\over c}\,\ve{n}\,\times\,(\dot{\ve{x}}_\o\,\times\,\ve{n})
\nonumber\\
&+&{1\over c^2}\,\Biggl\{\,(\ve{n}\,\cdot\,\dot{\ve{x}}_\o)\
\ve{n}\,\times\,(\dot{\ve{x}}_\o\,\times\,\ve{n})
+{1\over 2}\,\dot{\ve{x}}_\o\,\times\,(\ve{n}\,\times\,\dot{\ve{x}}_\o)
\,\Biggr\}
\nonumber\\
&+&{1\over c^3}\,\Biggl\{
\left({(\ve{n}\,\cdot\,\dot{\ve{x}}_\o)}^2+(1+\gamma)\,w(\ve{x}_\o)\right)\
\ve{n}\,\times\,(\dot{\ve{x}}_\o\,\times\,\ve{n})
+{1\over 2}\,(\ve{n}\,\cdot\,\dot{\ve{x}}_\o)\
\dot{\ve{x}}_\o\,\times\,(\ve{n}\,\times\,\dot{\ve{x}}_\o)
\Biggr\}
\nonumber\\
&+&\OO4,
\end{eqnarray}

\noindent
Here $\ve{s}=s^{a}$, $\ve{n}=n^i$, and for any $a^i$ and $b^i$
the Euclidean scalar and vector products are denoted as
$\ve{a}\,\cdot\,\ve{b}=\delta_{ij}\,a^i\,b^j=a^i\,b^i$ and
$\left(\ve{a}\times\ve{b}\right)^i=\varepsilon_{ijk}\,a^j\,b^k$,
respectively. Eq.
(\ref{aberration}) coincides with Eq. (7) of \cite{Klioner:2003}. The
discussion of the Lorentz transformations of the tetrads ${\rm
e}^{(\alpha)}_\mu$ at the end of Section \ref{Section-PPN-CoMRS-tetrad}
allows one to conclude that Eq. (\ref{aberration}) can be re-written as
a closed-form Lorentz transformation (see Section 5 of
\cite{Klioner:2003} for further details).

\section{Relativistic modeling of the rotational motion of the
satellite in the CoMRS}
\label{Section-rotational-motion}

In principle, one can consider the rotational motion of the satellite
relative to the CoMRS. To this end, the post-Newtonian equations of rotational
motion of a satellite relative to the CoMRS are necessary.
These equations have been
derived in \cite{Damour:Soffel:Xu:1993} in the framework of general
relativity and then extended to the PPN formalism in
\cite{Klioner:Soffel:1998b,Klioner:Soffel:2000}.
The final multipole-expanded form of the equations is
given in Section IX.F of \cite{Klioner:Soffel:2000}.
These are
differential equations for the post-Newtonian spin (total
angular momentum) $\S^a$ of the satellite:

\begin{equation}
\label{PPN-rotational-eqm}
\dot{\S}^a=\L^a,
\end{equation}

\noindent
where $\L^a$ is the post-Newtonian external torque, that can be
computed from the mechanical properties of the satellite
(inertial moments, etc.) and the ephemeris data of the Solar system
bodies (their BCRS positions, velocities, etc).

For a satellite on a heliocentric
orbit with the semi-major axes close to that of the Earth orbit, the
largest relativistic effect in its rotational motion relative to
 the kinematically non-rotating CoMRS is
clearly the geodetic precession which is of the order of $\sim
2\arcsec/{\rm cty}\approx2$~\muas/hr.

\section{Attitude description of the GAIA satellite}
\label{Section-attitude}

It is, however, clear that these dynamical equations of rotational
motion (at least the tiny relativistic corrections) play no
role in the accurate attitude determination of the satellite. As it is
discussed, e.g., in \cite{GAIA:2000} the attitude of the satellite
will be determined together with the astrometric parameters of the
sources from aposteriori processing of the observational data. The
attitude parameters are the parameters of the rotational matrix $P^a_{\ b}$
relating the CoMRS spatial axes $\X^a$ to the spatial axes
$\overline{\X}^a$ of the reference system in which the satellite's body
is fixed (the latter reference system is called Scanning Reference
System (SRS) in \cite{Bastian:2003}):

\begin{equation}
\label{SRS-X}
\overline{\X}^a=P^a_{\ b}\,\X^b.
\end{equation}

\noindent
The directly observable quantities for the missions like GAIA
(i.e. for scanning astrometric satellites) are the
coordinates $\overline{s}^a$
of the sources in the SRS tagged with the corresponding
time of observation in the satellite's proper time

\begin{equation}
\label{SRS-s}
\overline{s}^a=P^a_{\ b}\,s^b,
\end{equation}

\noindent
where $s^a$ is the vector defined by (\ref{s-(a)})--(\ref{s-a}). The
observables $\overline{s}^a$ should be first transformed from the SRS
into the CoMRS with the aid of matrix $P^a_{\ b}$ and then the
relativistic model as described e.g. in \cite{Klioner:2003} should be
applied to get the catalog positions of the sources. The matrix $P^a_{\
b}$ is clearly time-dependent and should be determined from the same
observations (a rough estimate of the matrix is provided by an apriori
dynamical modeling of the satellite rotation). The matrix $P^a_{\ b}$
can be parametrized with some Euler-like angles or in any other
suitable way.

In principle, any orientation of the CoMRS spatial axes can be used to
accomplish this data processing scheme and thus determine both the
astrometric parameters of the sources in the BCRS and the orientation
of the satellites's body relative to the CoMRS. However, the
kinematically non-rotating CoMRS represents a natural choice of the
orientation of the local coordinates.
Indeed, in this case the difference between the CoMRS
positions of the sources and the
catalogue positions comes from a number of well-understood effects
like proper motion, parallax, light deflection (all calculated in the
BCRS) and aberration (calculated as discussed in Section
\ref{Section-aberration} above). One can also argue that Eq.
(\ref{aberration}) takes its simplest form for the kinematically
non-rotating CoMRS: for any other spatial orientation of the CoMRS the
formula relating the observed direction to the source $\ve{s}$ to the
BCRS direction $\ve{n}$ differs from (\ref{aberration}) by
additional spatial rotation which exactly vanishes for the
kinematically non-rotating CoMRS.

One could in principle construct the tetrad directly for the SRS

\begin{eqnarray}
\label{tetrad-srs}
\overline{\rm e}^{(0)}_\mu&=&{\rm e}^{(0)}_\mu,
\nonumber\\
\overline{\rm e}^{(a)}_\mu&=&P^a_{\ b}\,{\rm e}^{(b)}_\mu,
\end{eqnarray}

\noindent
where ${\rm e}^{(\alpha)}_\mu$ is the tetrad of the CoMRS defined by
(\ref{monad-bcrs})--(\ref{triad-bcrsa}). This kind of tetrads was
discussed in \cite{Bini:deFelice:2003,Bini:Crosta:deFelice:2003}. The
tetrad $\overline{\rm e}^{(\alpha)}_\mu$ can be used to compute the
observed light direction $\overline{s}^a$ in the same way as described
in Section \ref{Section-aberration}. This way is totally equivalent to
using first the CoMRS tetrad to compute $s^a$ and then converting $s^a$
into $\overline{s}^a$ by using (\ref{SRS-s}). It is clear from
(\ref{tetrad-srs}) and, e.g., (\ref{s-(a)}) since the contraction is
associative and in particular $P^a_{\ b}\,{\rm
e}^{(b)}_\mu\,dx^\mu=(P^a_{\ b}\,{\rm e}^{(b)}_\mu)\,dx^\mu= P^a_{\
b}\,({\rm e}^{(b)}_\mu\,dx^\mu)$. In our opinion, however, keeping the
matrix $P^a_{\ b}$ in single formula (\ref{SRS-s}) allows one to
separate and simplify the relativistic and attitude models.

\section{Conclusions}
\label{Section-Conclusions}

The IAU 2000 relativity framework
\cite{IAU:2001,Rickman:2001,Soffel:et:al:2003} provides not only a
reasonable barycentric reference system for the whole solar system
(BCRS) and a physically adequate geocentric reference system for the
Earth neighborhood (GCRS). The IAU framework provides also suitable
tools to model any kind of astronomical observations. As it was
explicitly demonstrated above the same technique as used to construct
the GCRS can be applied to define a physically adequate proper
reference system of a test observer called CoMRS above. That reference
system can be used to model observation processes of any kind. The
coordinate basis of the CoMRS at its origin coincides with a particular
form of tetrad comoving with the observer. This means that the CoMRS
description of observables coincides with the classical tetrad
representation in cases where a tetrad is sufficient for modeling. On
the other hand, the CoMRS, being a well-defined 4-dimensional reference
system, is much more that just a tetrad and can be used to model
physical phenomena where spatial extension of the observer plays a role
(e.g., its rotational motion). In principle, any spatial orientation of
the CoMRS is possible. However, to model astronomical observations it
is preferable to adopt the kinematically non-rotating CoMRS having no
spatial rotation relative to the BCRS. This choice implies simplest
possible models for observables. The attitude of the observer can be
then described by a 3-dimensional spatial rotation in the CoMRS. This
allows one to separate the relativistic model and the attitude model
and to simply both of them as much as possible.

\end{document}